\documentclass[12pt,a4paper]{article}
\usepackage{amsmath}
\usepackage{amssymb}
\usepackage{enumerate}
\usepackage{amsfonts}
\usepackage[latin1]{inputenc}
\begin{document}

%
\newcounter{saveeqn}
\newcommand{\alpheqn}{\setcounter{saveeqn}{\value{equation}}%
\setcounter{equation}{0}%
\addtocounter{saveeqn}{1}
\renewcommand{\theequation}{\mbox{\arabic{saveeqn}\alph{equation}}}}
\newcommand{\reseteqn}{\setcounter{equation}{\value{saveeqn}}%
\renewcommand{\theequation}{\arabic{equation}}}
\begin{center}
{\bf ``Light-Front Quantization of the Polyakov D1 Brane Action \\ With a Scalar Dilaton Field'' \footnote{ ``Invited Contributed Talk'' at `` Workshop on Light-Cone 2007 Relativistic Hadronic and Nuclear Physics'', Ohio Center for Technology and Science,  Ohio State University, Columbus, Ohio, USA, May 14-18, 2007.}}

\vspace{14 mm}

D.S. Kulshreshtha $^{[{\rm a}]}$
\vspace{7 mm}

\begin{tabular}{cc}
a & Department of Physics and Astrophysics,\\
&  University of Delhi, Delhi, India. \\
& $<$dskulsh@gmail.com$>$
\end{tabular} 
\end{center}

\vspace{1cm}

\begin{abstract}
Light-front quantization of the conformally gauge-fixed Polyakov D1 brane action with a scalar dilaton field is studied using the equal light-cone world-sheet time framework. 
\end{abstract}

\section{\bf Polyakov D1 Brane Action With a Scalar Dilaton Field in the Conformal Gauge}

In the present work, we study the light-front(LF) quantization of the Polyakov D1 brane action\cite{1,2} with a scalar dilaton field  in the conformal gauge (CG), in the front-frame(FF) of dynamics using the equal light-cone(LC) world-sheet (WS) time framework on the hyperplanes of the LF defined by the LC-WS time ${\sigma}^{+} = (\tau + \sigma) =$ constant. The Polyakov D1 brane action describes the propagation of a D1 brane in a d-dimensional curved background $h_{\alpha \beta}$ (with $ d = 10$ for the fermionic and $ d = 26$ for bosonic D1 brane) defined by \cite{1,2}:
\alpheqn
\begin{eqnarray}
{\tilde S} &=& \int {\tilde{\cal L}} d^{2} \sigma \\
\label{1a}
{\tilde{\cal L}}  &=&  \biggl[ - \frac{T}{2} [ e^{-\varphi}] {\sqrt{-h}} h^{\alpha\beta} G_{\alpha\beta} \biggr] \quad ; \quad  h = \det (h_{\alpha\beta}) \\
\label{1b}
G_{\alpha\beta} &=& {\partial}_{\alpha} X^{\mu} {\partial}_{\beta} X^{\nu} {\eta}_{\mu\nu}; ~ {\eta}_{\mu\nu} = {\rm diag} (-1,+1,...,+1) \\
\label{1c}
\mu,\nu &=& 0,1,i \quad;\quad  i = 2,3,....,(d-1)\quad;\quad \alpha,\beta = 0,1
\label{1d} 
\end{eqnarray}
\reseteqn
Here $ {\sigma}^{\alpha} \equiv (\tau,\sigma)$ are the two parameters describing the WS. $T$ is the string tension and the overdots and primes would denote the derivatives with respect to $\tau$ and $\sigma$.  $G_{\alpha\beta}$ is the induced metric on the WS and $X^{\mu}(\tau, \sigma)$ are the maps of the WS into the $d$-dimensional Minkowski space and describe the strings evolution in space-time \cite{1,2}. $h_{\alpha\beta}$ are the auxiliary fields (which turn out to be proportional to the metric tensor ${\eta}_{\alpha\beta}$ of the two-dimensional surface swept out by the string). One can think of ${\tilde S} $ as the action describing $d$ massless scalar fields $X^{\mu}$ in two dimensions moving on a curved background $h_{\alpha\beta}$. Also because the metric components $h_{\alpha\beta}$ are varied in the above equation, the 2-dimensional gravitational field $h_{\alpha\beta}$ is treated not as a given background field, but rather as an adjustable quantity coupled to the scalar fields \cite{1}. The action ${\tilde S} $ has the well-known three local gauge symmetries given by the 2-dimensional WS reparametrization invariance (WSRI) and the Weyl invariance (WI) \cite{1,2}. We could  use the three local gauge symmeties of the theory to choose $h_{\alpha\beta}$ to be of a particular form \cite{1,2}:
\begin{equation}
h_{\alpha\beta} = {\eta}_{\alpha\beta} = 
\left( \begin{array}{ll} -1 & ~~0 \\ ~~0 & +1 \end{array} \right) \quad ;\quad 
{\sqrt{-h}} = {\sqrt{-\det(h_{\alpha\beta})} } = +1
\label{2}
\end{equation} 
This is the so-called conformal gauge (CG) and the action ${\tilde S} $ in this CG becomes:
\alpheqn
\begin{eqnarray}
S_{1}  &=&  \int {\cal L}_{1}  d^{2} \sigma \\
\label{3a}
{\cal L}_{1}  &=& (-T/2)[ e^{-\varphi}] {\sqrt{-h}} h^{\alpha\beta} G_{\alpha\beta}  \\
\label{3b}
 &=&  (-T/2)[ e^{-\varphi}] {\partial}^{\beta} X^{\mu} {\partial}_{\beta} X_{\mu} \\
\label{3c}
\mu,\nu &=& 0,1,i \quad;\quad  i = 2,3,....,(d-1) \quad;\quad  
\alpha,\beta = 0,1
\label{3d} 
\end{eqnarray} 
\reseteqn

\section{Light-Front Quantization (LFQ) }

For the LFQ of the theory we use the three local gauge symmetries of the theory to choose $h_{\alpha\beta}$ to be of a particular form as follows:
\alpheqn
\begin{equation}
h_{\alpha\beta} := {\eta}_{\alpha\beta} = 
\left( \begin{array}{ll}  ~~~~0 & -1/2 \\ -1/2 & ~~~~0 \end{array} \right) 
\quad;\quad
{\sqrt{-h}} = {\sqrt{- \det(h_{\alpha\beta)}} } = + 1/2
\label{4a}
\end{equation} 
and
\begin{equation}
h^{\alpha\beta} := {\eta}^{\alpha\beta} = \left( \begin{array}{ll}  ~~0 & -2 \\ -2 & ~~0 \end{array} \right)
\label{4b}
\end{equation}
\reseteqn 
This is the so-called conformal gauge (CG) in the LFQ of the theory. Also, in this formulation, we use the LC variables defined by \cite{1,2}:
\begin{equation}
{\sigma}^{\pm} := (\tau \pm \sigma) \quad {\rm and} \quad X^{\pm} := (X^{0} \pm X^{1} )/ {\sqrt{2}}
\label{5}
\end{equation}

The conformally gauge-fixed Polyakov D1 brane LC action in a $d$-dimensional flat background $h_{\alpha\beta}$ in the presence of a scalar dilation field $\varphi$ is defined by \cite{1,2}:
\alpheqn
\begin{eqnarray}
S_{2} &=& \int {\cal L}_{2} d {\sigma}^{+} d {\sigma}^{-} \\
\label{6a}
{\cal L}_{2} &=& [ e^{-\varphi}{\cal L}_{1}]\\
\label{6b}
&=& \biggl[ - \frac{T}{2}  e^{- \varphi}\biggr] \biggl[ {\partial}^{\beta} X^{\mu} {\partial}_{\beta} X_{\mu} \biggr]   \\ 
\label{6c}
&=& \biggl[ -\frac{T}{2} e^{-\varphi} \biggr] \biggl[ ( {\partial}_{+} X^{+} ) ( {\partial}_{-} X^{-} ) + ( {\partial}_{+} X^{-} ) ({\partial}_{-} X^{+} ) + ( {\partial}_{+} X^{i}) ( {\partial}_{-} X^{i} )\biggr] \\
\label{6d}
& & \mu, \nu = +,-,i  \quad ;\quad  i = 2,3,...(d-1) \quad;\quad 
\alpha,\beta = +,-
\label{6e}
\end{eqnarray}
\reseteqn
where $d = 10$ for the fermionic string and $d = 26$ for the bosonic string. In the present work we consider only the bosonic case. In the following we study the LC action $S_{2}$  which describes the Polyakov D1 brane LC action. The theory is seen to possess 27 primary constraints:
\alpheqn
\begin{eqnarray}
{\Omega}_{1} &=& \pi \approx 0 \\
\label{7a}  
{\Omega}_{2} &=& (P^{+} + \frac{T}{2} e^{-\varphi}( {\partial}_{-} X^{+} ) 
\approx 0 \\
\label{7b}
{\Omega}_{3} &=& ( P^{-} + \frac{T}{2} e^{-\varphi}( {\partial}_{-} X^{-} ) 
\approx 0 \\ 
\label{7c}
{\Omega}_{i} &=&  ( P_{i} + \frac{T}{2} e^{-\varphi}( {\partial}_{-} X^{i} ) \quad;\quad i =  2,3,.......(d-1).
\approx 0
\label{7d}
\end{eqnarray}
\reseteqn
 Where $ \pi, P^{+}, P^{-}$ and $P_{i}$ are the canonical momenta conjugate respectively to $\varphi, X^{-},X^{+} $, and $X^{i}$. After including these 27 primary constraints in the canonical Hamiltonian density ${\cal H}^{c}_{2}$ with the help of Lagrange multiplier fields $u,v,w$ and $z_{i}$, the total Hamiltonian density ${\cal H}^{T}_{2}$ could be written as
\begin{eqnarray}
{\cal H}^{T}_{2} = \biggl[ u \pi + v ( P^{+} + \frac{T}{2} e^{-\varphi} {\partial}_{-} X^{+}) + w(P^{-} + \frac{T}{2} e^{-\varphi} {\partial}_{-} X^{-} ) 
 + z_{i}(P_{i} + \frac{T}{2}e^{-\varphi} {\partial}_{-} X^{i}) \biggr]
\label{8}
\end{eqnarray}
We treat $u, v,,w$ and $z_{i}$ as dynamical. Demanding that the primary constraints  be preserved in the course of time one does not get any secondary constraints. The theory is thus seen to possess only 27 constraints ${\Omega}_{1},~ {\Omega}_{2},~ {\Omega}_{3}$ and ${\Omega}_{i}$. The Hamiltons equations obtained from the total Hamiltonian $ H^{T}_{2} $  describe the correct dynamics of the system. Now following the standard Dirac qunatization procedure in the Hamiltonian formulation \cite{4}, the nonvanishing ELCWST Dirac brackets of the theory described by the Polyakov D1 brane LC action $S_{2}$ in the presence of the scalar dilation field $\varphi$ could be obtained after a lengthy but straightforward calculation and are omitted here for the sake of brevity\cite{3,4,5}. In the path integral formulation, the transition to quantum theory is made by writing the generating fucntional $Z_{2} [J_{k}]$ for the theory in the presence of the external sources $J_{k}$  which is  obtained for the present theory as \cite{2,3}: 
\begin{equation}
Z_{2}[J_{k}] := \int [d\mu] \exp \biggl[ i \int d {\sigma}^{+} d {\sigma}^{-} 
 [ (- T/2) e^{- \varphi} ] [ v ( {\partial}_{-} X^{+} ) + w ({\partial}_{-} X^{-} ) + z_{i} ( {\partial}_{-} X^{i} ) ] + J_{k} {\Phi}^{k} \biggr] 
\label{9}
\end{equation}
where the phase space variables of the theory defined by the action $S_{2}$ are ~$ {\Phi}^{k} \equiv (\varphi , X^{+}, X^{-},X^{i}, u, v, w, z_{i})$ with the corresponding respective canonical conjugate momenta: ${\Pi}_{k} \equiv (\pi,P^{-},P^{+},P^{i},p_{u},p_{v},p_{w},p_{z_{i}})$. The functional measure $[d\mu]$ of the generating functional $Z_{2}[J_{k}]$ is finally obtained as\cite{2,3}:
\begin{eqnarray}
[d\mu] &=& \biggl[ \frac{1}{2} R T^{2} e^{- 2\varphi} {\partial}_{-} \delta({\sigma}^{-} - {\sigma'}^{-} ) \biggr] [d\varphi][dX^{+}][dX^{-}][dX^{i}][du][dv][dw][dz_{i} ] \nonumber \\
& & [d\pi][dP^{-}][dP^{+}][dP_{i}][dp_{u}][dp_{v}][dp_{w}][dp_{z_{i}}] \delta [(P_{i} + \frac{T}{2}  e^{-\varphi} ( {\partial}_{-} X^{i}) \approx 0] \nonumber \\
& & \delta [P^{+} + \frac{T}{2}  e^{-\varphi} ( {\partial}_{-} X^{+} ) \approx 0] \delta[P^{-} + \frac{T}{2}  e^{-\varphi} ( {\partial}_{-} X^{-}) \approx 0].
\label{10}
\end{eqnarray}

The LC Hamiltonian and path integral quantization of the Polyakov D1 brane LC action $S_{2}$ in the presence of the scalar dilaton field $\varphi$  under the conformal gauge, using the ELCWST framework on the hyperplanes of the world-sheet defined by the LC world-sheet time: $\sigma^{+} = (\sigma + \tau)$ = constant,  is now complete. Also because this is a (conformally) gauge-fixed action, the theory is therefore gauge noninvariant as expected and the associated constraints of the theory form a set of second-class constraints. The problem of operator ordering that occurrs while making a transition from the Dirac brackets to the corresponding commutation relations could be resolved  by demanding that all the string fields and momenta of the theory are Hermitian operators and that all the canonical commutation relations be consistent with the hermiticity of these operators \cite{2}. It is important to mention here that in our work we have not imposed any boundary conditions (BC's) for the open and closed strings separately. There are two ways to take these BC's into account: (a) one way is to impose them directly in the usual way for the open and closed strings separately in an appropriate manner \cite{1,2}, and (b) an alternative second way is to treat these BC's as the Dirac primary constraints \cite{6} and study the theory accordingly \cite{6}.

\end{document}